\documentclass[conference]{IEEEtran}
\IEEEoverridecommandlockouts

\usepackage{amsmath, amssymb, amsthm, amsfonts}
\usepackage[ruled,vlined]{algorithm2e}
\usepackage[english]{babel}
\usepackage{listings}
\usepackage{tikz}
\usetikzlibrary{arrows.meta,positioning}
\theoremstyle{definition}
\newtheorem{definition}{Definition}[section]
\theoremstyle{remark}

\usepackage{graphicx}
\usepackage{url}
\usepackage{hyperref}
\usepackage{orcidlink}
\usepackage[capitalize]{cleveref}
\usepackage{cite}
\usepackage{amsmath,amssymb,amsfonts}
\usepackage{algorithmic}
\usepackage{textcomp}
\usepackage{xcolor}
\usepackage{quantikz}
\usepackage{amsmath, amssymb, amsthm, amsfonts}
\usepackage[english]{babel}
\usepackage{listings}
\usepackage{tikz}
\usetikzlibrary{arrows.meta,positioning}

\usepackage{subcaption}

\usepackage[ruled,vlined]{algorithm2e}
\SetKwComment{tcp}{\% }{}
\SetKwFunction{InitializeTN}{InitializeTN}
\SetKwFunction{InitializeState}{InitializeState}
\SetKwFunction{SyncKetFromBra}{SyncKetFromBra}
\SetKwFunction{ConditionHolds}{ConditionHolds}
\SetKwFunction{CompressionStep}{CompressionStep}
\SetKwFunction{MeasureProb}{MeasureProb}
\SetKwFunction{Collapse}{Collapse}
\SetKwFunction{TopKByWeight}{TopKByWeight}
\SetKwFunction{DeepCopy}{DeepCopy}
\SetKwFunction{Prod}{Prod}
\crefname{section}{Sec.}{Secs.}
\Crefname{section}{Sec.}{Secs.}
\crefname{subsection}{Sec.}{Secs.}
\Crefname{subsection}{Sec.}{Secs.}
\crefname{subsubsection}{Sec.}{Secs.}
\Crefname{subsubsection}{Sec.}{Secs.}
\Crefname{figure}{Fig.}{Figs.}
\Crefname{table}{Tab.}{Tabs.}
\crefname{definition}{Def.}{Defs.}
\Crefname{definition}{Def.}{Defs.}

\SetAlFnt{\small}

\SetCommentSty{mycommfont}

\def\BibTeX{{\rm B\kern-.05em{\sc i\kern-.025em b}\kern-.08em
    T\kern-.1667em\lower.7ex\hbox{E}\kern-.125emX}}
\begin{document}

\title{Tensor Network Simulation of Dynamic Circuits\\
}

\author{
\IEEEauthorblockN{Innocenzo Fulginiti \orcidlink{0000-0001-8818-9626}}
\IEEEauthorblockA{\textit{CIT, Department of Computer Science} \\
\textit{Technical University of Munich}\\
85748 Garching, Germany \\
innocenzo.fulginiti@tum.de}
\and
\IEEEauthorblockN{Alessandro Poggiali \orcidlink{0000-0002-1591-7925}}
\IEEEauthorblockA{\textit{Department of Computer Science} \\
\textit{University of Pisa}\\
Pisa, Italy\\
alessandro.poggiali@unipi.it}
\and
\IEEEauthorblockN{Christian B.~Mendl \orcidlink{0000-0002-6386-0230}}
\IEEEauthorblockA{\textit{CIT, Department of Computer Science} \\
\textit{Technical University of Munich}\\
85748 Garching, Germany \\
christian.mendl@tum.de}
}

\newcommand{\ketfixed}[1]{\lvert #1 \rangle}
\newcommand{\brafixed}[1]{\langle #1 \rvert}
\newcommand{\braketfixed}[2]{\langle #1 \mid #2 \rangle}

\maketitle

\begin{abstract}
Dynamic circuits, which incorporate mid-circuit measurements and classically controlled operations, extend the expressive power of quantum programs and are central to applications such as quantum error correction, state preparation, and distributed quantum algorithms. However, their classical simulation is challenging due to the exponential growth of execution branches induced by mid-circuit measurements.

In this work, we develop a tensor network approach for simulating dynamic circuits by extending a DMRG-based method for quantum circuit simulation to support dynamic circuit operations. To address the exponential proliferation of execution branches, we introduce and compare two strategies: a single-path stochastic approach that samples individual measurement outcomes, and a multi-path approach that maintains an ensemble of branches. We benchmark these methods on standard dynamic circuits, including the teleportation protocol and GHZ state preparation, as well as on random dynamic circuits. Our analysis explores the trade-offs between simulation accuracy and computational efficiency. These results provide a practical and scalable framework for the classical simulation of dynamic circuits.
\end{abstract}


\section{Introduction}
\label{sec:introduction}
Quantum computing holds the promise of transforming computation, offering a path to solving problems that remain fundamentally out of reach for classical hardware. However, current machines are still far from delivering on these promises, since noise, decoherence, and gate errors remain fundamental obstacles, limiting circuit depth and undermining the reliability of current devices. In this scenario, efficient classical simulation of quantum systems plays a fundamental role in validating and benchmarking quantum hardware and in developing deeper insight into the behavior of quantum algorithms. 
Among the most practically significant advances in quantum circuit design is the emergence of dynamic circuits, a class of circuits that incorporate mid-circuit measurements, allowing subsequent gates to be conditioned on measurement outcomes. This capacity for classical feedback within the circuit itself substantially expands the expressive power of quantum computation beyond what static circuit architectures can achieve. Dynamic circuits are a cornerstone of quantum error correction \cite{majumder2020real, ryan2021realization, botelho2022error}.
They enable efficient state preparation protocols that would otherwise require deeper circuits \cite{smith2024constant, baumer2025measurement, buhrman2024state}, as well as reducing the preparation cost of well-known states like GHZ states \cite{baumer2024efficient}.
Moreover, dynamic circuits arise naturally in quantum communication protocols, where real-time classical processing must be interleaved with quantum evolution \cite{bennett1993teleporting, carrera2024combining}. Their broad relevance across these settings makes the accurate and scalable simulation of dynamic circuits an increasingly important challenge.

In this work, we study the classical simulation of dynamic circuits within the framework of tensor networks, which are particularly well suited for representing and manipulating quantum states. Tensor network techniques constitute a central paradigm in classical simulation, supporting state-of-the-art methods in condensed matter physics, quantum chemistry, and quantum information, and are employed for benchmarking and simulating quantum circuits \cite{orus2014practical, markov2008simulating, pan2022simulation, seitz2023simulating}. Specifically, we extend the tensor network simulation framework based on the Density Matrix Renormalization Group (DMRG) algorithm \cite{white1993density, schollwock2011density}, recently used for static unitary circuits for Matrix Product States (MPS) \cite{ayral2023simulationmps}, and for Tree Tensor Networks (TTN) \cite{dubey2025simulating} topologies, to natively incorporate mid-circuit measurements, enabling the tracking of quantum states across measurement-conditioned branches. A key challenge in this setting is that the number of distinct execution branches grows exponentially with the number of intermediate measurements, rendering an exact enumeration of all branches computationally infeasible at scale. To address this, we introduce and benchmark single-path stochastic sampling of measurement outcomes alongside a multi-path approach that retains a weighted ensemble of branches. Our method enables the simulation of quantum states using both MPS and TTN. 
We validate our approach on standard dynamic-circuit benchmarks, including quantum teleportation and GHZ-state preparation, and systematically compare single-path and multi-path strategies on randomly generated dynamic circuits.

\section{Related Works}
\label{sec:related_works}

Recent quantum hardware platforms support dynamic circuits, extending static circuits with mid-circuit measurements and classical feedforward \cite{corcoles2021exploiting, ryan2017hardware, decross2023qubit}. We distinguish these from non-adaptive circuits, where measurements occur but do not affect subsequent gates, allowing simulation via trajectory sampling without branching \cite{czischek2021simulating, doggen2022generalized, yanay2024detecting}. Dynamic circuits, by contrast, use measurement outcomes to determine later gates, introducing classical control flow. This capability is central to a broad range of applications, including real-time feedback and error mitigation on quantum hardware \cite{majumder2020real, ryan2021realization, botelho2022error}, efficient state preparation \cite{smith2024constant, baumer2025measurement, buhrman2024state, baumer2024efficient}, and distributed quantum algorithms \cite{bennett1993teleporting, carrera2024combining}. Dynamic circuits have been explored experimentally and from a compilation perspective. In \cite{koh2022experimental}, simulations of random circuits with mid-circuit measurements are used to predict the volume-law to area-law entanglement transition later observed experimentally. The works in  \cite{fulginiti2024reducing, fulginiti2025optimization, fulginiti2026}, propose a probabilistic circuit model which can be exploited to optimize dynamic circuit by replacing certain mid-circuit measurements and resets with randomized unitary gates, while in \cite{fulginiti2026bqcp}, a compile-time pass for simplifying dynamic circuits is proposed. 
From a tensor network perspective, \cite{pereira2025efficient} simulates circuits with repeated mid-circuit weak measurements by encoding the circuit as a tensor network and sampling outcomes via a Markov-chain process, avoiding explicit enumeration of all branches. However, it is restricted to non-adaptive circuits, where outcomes do not affect subsequent gates, and does not extend to dynamic, measurement-conditioned circuits.
A complementary work \cite{deliyannis2022improving} uses MPS simulators to validate dynamic circuits with mid-circuit measurements and classical control, but treats MPS as a black-box backend rather than tackling dynamic circuit simulation generally.
In this work, we focus on DMRG-inspired tensor network methods, which have recently been extended to simulate general static unitary circuits \cite{ayral2023simulationmps, dubey2025simulating}. Despite this progress, incorporating both mid-circuit measurements and classical feedforward remains an open challenge. We address this gap by developing a DMRG-based framework that explicitly captures the branching structure induced by classical feedforward.

\section{Preliminaries}
\label{sec:preliminaries}
In this section, we recall the basic notions used throughout the paper. We describe dynamic circuits with mid-circuit measurements and classically controlled operations (\cref{sec:prel-quantumstates-dyncirc}). We review the representation of pure quantum states using tensor networks (\cref{sec:prel-tn-rep-qs}), and we describe their use for simulating quantum circuits via the DMRG algorithm (\cref{subsec:prelim-sim-dmrg}).
\subsection{Quantum states, quantum operations and dynamic circuits}
\label{sec:prel-quantumstates-dyncirc}
\begin{definition}[Pure quantum state]
\label{def:pure_state}
A pure quantum state on $n$ qubits is a normalized vector in $\mathbb{C}^{2^n}$ of the form
\(
\ket{\psi}
=
\sum_{q_1,\dots,q_n \in \{0,1\}}
\psi_{q_1\dots q_n}
\ket{q_1\dots q_n},
\)
where each $q_i \in \{0,1\}$ labels the state of qubit $i$, the vectors $\ket{q_1\dots q_n}$ form the computational basis, and the coefficients
$\psi_{q_1\dots q_n} \in \mathbb{C}$ are the corresponding amplitudes, such that
$\sum_{q_1,\dots,q_n \in \{0,1\}} \left|\psi_{q_1\dots q_n}\right|^2 = 1.$
\end{definition}


\begin{definition}[Unitary operation]
\label{def:unitary_operation}
A unitary operation acting on $k$ qubits is a linear operator
$U : (\mathbb{C}^2)^{\otimes k} \to (\mathbb{C}^2)^{\otimes k}$ satisfying
$U^\dagger U = I$.
Its application to a $k$-qubit pure state $\ket{\psi}$ yields another pure state $\ket{\psi'} = U\ket{\psi}$.
If $U$ acts on a subset of $k$ qubits within an $n$-qubit system, it is implicitly embedded in the full system by tensoring it with identity operators on the remaining qubits.
\end{definition}

\begin{definition}[Mid-circuit measurement]
\label{def:mid_circuit_measurement}
A mid-circuit measurement is a non-unitary operation that measures a qubit during the execution of the circuit and produces a classical outcome.
For a measurement in the computational basis, the possible outcomes are $m \in \{0,1\}$, occurring with probability
\(
p_m = \bra{\psi} P_m \ket{\psi},
\)
where $P_m$ is the projector associated with outcome $m$.
Conditioned on observing outcome $m$, the post-measurement state is
\(
\ket{\psi_m} = \frac{P_m\ket{\psi}}{\sqrt{p_m}},\) with probability \(p_m>0.
\)
\end{definition}

\begin{definition}[Classically controlled operation]\label{def:classically_controlled_operation}
A classically controlled operation is a unitary operation $U$ whose application depends on a Boolean condition on classical bits: the unitary $U$ is applied during the execution of the circuit if the condition is satisfied; otherwise, it is skipped.
\end{definition}

\begin{definition}[Dynamic circuit]
\label{def:dynamic_quantum_circuit}
Let $Q=\{Q_1,\dots,Q_n\}$ be a quantum register of $n$ qubits, and let $C=\{C_1,\dots,C_m\}$ be a classical register of $m$ bits. Let \(O = (O_1,\dots,O_s)\) be an ordered sequence of operations, where each $O_i$ is either a unitary operation acting on qubits in $Q$, a mid-circuit measurement of a qubit in $Q$ whose outcome is stored in a bit of $C$, or a classically controlled operation whose condition is evaluated on the bits of $C$.
A dynamic circuit $D$ with quantum register $Q$, classical register $C$, and sequence of operations $O$ is the triple
\(
D = ( Q, C, O ).
\)
\end{definition}

The pure-state representation of \cref{def:pure_state} is sufficient for unitary circuits, i.e., circuits containing only unitary operations. In contrast, dynamic circuits include mid-circuit measurements that produce random outcomes and induce different conditional quantum-classical evolutions, and their states can be described as a collection of branches.

\begin{definition}[State of a dynamic circuit]
\label{def:dynamic_circuit_state}
The state of a dynamic circuit is represented as a collection of branches
\(
\mathcal{B} = \{(p_\gamma, c_\gamma, \ket{\psi_\gamma})_\gamma\},
\)
where for each branch $(p_\gamma, c_\gamma, \ket{\psi_\gamma})_\gamma \in \mathcal{B}$, \(c_\gamma \in \{0,1\}^m\) is the classical-register configuration of the branch, \(\ket{\psi_{\gamma}}\) is the corresponding pure quantum-register state, and $p_\gamma \geq 0$, with $\sum_\gamma p_\gamma = 1$, represents the probability to reach the classical-quantum state configuration $(c_\gamma, \ket{\psi_\gamma})$.

\end{definition}


\subsection{Representing pure quantum states with tensor networks}
\label{sec:prel-tn-rep-qs}
Consider an $n$-qubit pure state
\(
\ket{\psi}
=
\sum_{q_1,\dots,q_n \in \{0,1\}}
\psi_{q_1\dots q_n}
\ket{q_1\dots q_n}
\). The amplitudes $\psi_{q_1\dots q_n}$ form an order-$n$ tensor, with one index of dimension $2$ associated with each qubit. A direct representation of $\ket{\psi}$ requires storing $2^n$ complex coefficients, therefore, storing the full state requires resources that grow exponentially with the number of qubits of the system. 
Tensor networks address this limitation by factorizing the full tensor of amplitudes into a collection of smaller tensors connected by auxiliary indices. Instead of storing the coefficient tensor explicitly, one represents it as the contraction of local tensors, each carrying one or more physical indices associated with qubits together with additional bond (or virtual) indices connecting neighboring tensors.
Formally, the amplitudes can be written as
\(
\psi_{q_1\dots q_n}
=
\sum_{\{\alpha\}}
\prod_{v\in V}
T^{[v]}_{q(v),\,\alpha(v)},
\)
where $V$ is the set of tensors in the network, $q(v)$ denotes the collection of physical indices attached to the tensor $v$, $\alpha(v)$ denotes the bond indices attached to the tensor $v$, and $\{\alpha\}$ denotes the collection of all internal bond indices of the tensor network. The summation runs over all possible values of these internal bond indices. Equivalently, the quantum state can be expressed as
\[
\ket{\psi}
=
\sum_{q_1,\dots,q_n \in \{0,1\}}
\left(
\sum_{\{\alpha\}}
\prod_{v\in V}
T^{[v]}_{q(v),\alpha(v)}
\right)
\ket{q_1\dots q_n}.
\]

The efficiency of this representation depends on the dimensions $\chi$ of the bond indices. When these remain moderate, the number of parameters required to describe the state grows much more slowly than the full Hilbert-space dimension $2^n$. Tensor networks therefore provide a compressed representation of quantum states by exploiting the structure of their correlations and entanglement. Since our method supports Matrix Product States (MPS) and Tree Tensor Networks (TTN), two widely used tensor network representations for many-body quantum states, we briefly recall their definitions.

\paragraph{Matrix Product States}
In an MPS representation, an \(n\)-qubit state is described by a chain of \(n\) local tensors \(T^{[0]},\dots,T^{[n-1]}\), one for each qubit, connected by \(n-1\) virtual bonds with bond dimensions \((\chi_1,\dots,\chi_{n-1})\). Each tensor carries one physical index of dimension \(2\). The boundary tensors have one virtual index, while the internal tensors have two, corresponding to their left and right neighbors.

\paragraph{Tree Tensor Networks}
In a TTN representation, an \(n\)-qubit state is described by a rooted hierarchical tensor network specified by a structure \((n_0,n_1,\dots,n_L)\), where \(n_l\) denotes the number of tensors in layer \(l\), with \(n_0=1\) at the root. In our implementation based on \cite{dubey2025simulating}, the final level \(L\) does not contain tensors: it consists only of the \(n=n_L\) open physical indices corresponding to the qubits. Accordingly, the tensors are located only in layers \(0,\dots,L-1\), and the tensors in the lowest tensor layer \(L-1\) are directly connected to the physical indices. We consider regular TTN hierarchies, in which each tensor in a given layer is connected to the same number of tensors in the layer below. Internal edges define the virtual indices and their associated bond dimensions \(\{\chi_e\}\).

\subsection{Simulating unitary quantum circuits with DMRG}
\label{subsec:prelim-sim-dmrg}
The Density Matrix Renormalization Group (DMRG) algorithm \cite{white1993density, schollwock2011density} is a variational method originally introduced for approximating ground states of one-dimensional quantum systems. It can be adapted to the simulation of quantum circuits by representing quantum states through tensor networks and quantum gates as local tensor operators \cite{ayral2023simulationmps, dubey2025simulating}.

Let $\ket{\psi^i}$ denote the quantum state of the circuit after the application of the first gates $i$ and let $\ketfixed{\tilde\psi^i}$ denote its tensor network approximation. 
\begin{definition}[Unitary chunk]
A unitary chunk of length $k$ starting at position $i+1$ is the ordered product of $k$ consecutive unitary gates
\(U_{i+k}\cdots U_{i+1}.\)
Applying the chunk to the exact state $\ket{\psi^i}$ yields
\(\ket{\psi^{i+k}} = U_{i+k}\cdots U_{i+1}\ket{\psi^i}.\)
\end{definition}

\begin{definition}[Compression step]
Given a tensor network approximation $\ketfixed{\tilde\psi^i}$ and a unitary chunk $U_{i+k}\cdots U_{i+1}$, a compression step is the variational procedure that computes a new tensor network state $\ketfixed{\tilde\psi^{i+k}}$ approximating the evolved state
\(\ket{\psi^{i+k}} \)
by maximizing the overlap
\(|\braketfixed{\tilde\psi^{i+k}}{\psi^{i+k}}|^2.\)
\end{definition}

The simulation is therefore organized as a sequence of compression steps, each associated with a unitary chunk. At each step, a chunk of $k$ gates is applied to the current approximate state, and the resulting evolved state is then variationally compressed within the chosen tensor network ansatz. The target state $\ket{\psi^{i+k}}$ is not constructed explicitly in the full Hilbert space; rather, it is represented implicitly as a contracted tensor network obtained by composing the tensors of $U_{i+k}\cdots U_{i+1}$ with those of $\ketfixed{\tilde\psi^i}$. This provides an efficient description whose cost is governed by the bond dimensions of the tensor network rather than by the dimension of the full Hilbert space.
Each compression step is performed variationally by optimizing one tensor at a time while keeping all the others fixed. For a given tensor $\tilde T^{[v]}$ of the ansatz $\ketfixed{\tilde\psi^{i+k}}$, the overlap to be maximized is represented as a tensor network obtained by connecting the ansatz tensors, the gates in the unitary chunk, and the tensors of the previous approximation $\ketfixed{\tilde\psi^i}$. Contracting all tensors in this tensor network except $\tilde T^{[v]*}$ yields an environment tensor $F^{[v]}$, reducing the overlap to a simple inner product $\braketfixed{\tilde \psi^{i+k}}{\psi^{i+k}}$. The local optimization problem, $\max_{T^{[v]}}\braketfixed{T^{[v]}}{F^{[v]}}$,
is then solved analytically, and the optimal tensor is given by
\(
\tilde T_{\text{opt}}^{[v]} = \frac{F^{[v]}}{\sqrt{\braket{F^{[v]}}{F^{[v]}}}},
\)
where $\braket{F^{[v]}}{F^{[v]}} = \sum_{i_1,\dots,i_r}
|F^{[v]}_{i_1\dots i_r}|^2$ denotes the full contraction of $F^{[v]}$ with its complex conjugate. The whole procedure is iterated over a fixed number of sweeps $s$, which is typically small and sufficient to achieve convergence to the global optimum \cite{ayral2023simulationmps}. For each unitary chunk, we define the partial fidelity
\(
f_{i\to i+k}
=
|\braketfixed{\tilde \psi^{i+k} | U_{i+k}\dots U_{i+1}}{\psi^i}|^2,
\)
which measures the quality of the approximated state during the simulation. The global fidelity is then estimated as
\(
\tilde{\mathcal F}
=
\prod_{(i\to i+k)} f_{i\to i+k},
\)
namely as the product of the partial fidelities associated with all compression steps.

\section{Method}
\label{sec:method}
In this section, we present our extension of the DMRG-based simulation workflow of \cite{ayral2023simulationmps}, recalled in \cref{subsec:prelim-sim-dmrg}, to the simulation of dynamic circuits.
We first describe how we represent the hybrid classical-quantum state associated with an execution branch (\cref{sec:method-state-representation}), and how dynamic operations are handled at the branch level (\cref{sec:method-dynamic-ops}). We then introduce the rule used to construct and process unitary chunks in the presence of dynamic operations (\cref{sec:method-chunk-dyn-circ}). Next, we present two simulation strategies for dynamic circuits: a single-path strategy, which propagates one execution branch per run, and a multi-path strategy, which propagates more active branches simultaneously, truncating low-probability branches when their number exceeds a prescribed threshold (\cref{sec:method-dynamic-workflow}). Finally, we analyze correctness and complexity of our method (\cref{sec:method-correctness-complexity}).

\subsection{Representation of the classical-quantum state}
\label{sec:method-state-representation}
Consider a dynamic circuit over a quantum register \(Q=\{Q_1,\dots,Q_n\}\) and a classical register \(C=\{C_1,\dots,C_m\}\).
Following the definition of state of a dynamic circuit given in \cref{def:dynamic_circuit_state}, we represent a single branch of the dynamic circuit state as
\(
(p_{\gamma}, c_{\gamma}, \ketfixed{\tilde\psi_{\gamma}}),
\)
where \(p_{\gamma} > 0\) is the branch probability, \(c_{\gamma} \in \{0,1\}^m\) is the configuration of the classical register $C$, and \(\ketfixed{\tilde\psi_{\gamma}}\) is the tensor-network approximation of the pure quantum state of the quantum register $Q$.
\subsubsection{Classical state}
The classical register state is represented as a $m$-bit vector \(
c_{\gamma}=(c_1,\dots,c_m)\in\{0,1\}^m,
\)
whose entries are updated whenever a mid-circuit measurement writes an outcome into a classical bit $C_i \in C$.
\subsubsection{Quantum state}
For the quantum register state $\ketfixed{\psi}_{\gamma}$ we use the tensor-network representations introduced in \cref{sec:prel-tn-rep-qs}, supporting MPS and TTN topologies. Each internal edge \(e\) of the tensor network carries an auxiliary bond index \(\alpha_e\), whose range is determined by the corresponding bond dimension \(\chi_e\). The complexity of the representation is therefore controlled by the bond dimensions \(\{\chi_e\}\).

\subsection{Handling dynamic operations}
\label{sec:method-dynamic-ops}

\subsubsection{Mid-circuit measurements}
\label{sec:method-dynamic-ops-mcm}
Consider an execution branch $(p_{\gamma}, c_{\gamma}, \ketfixed{\tilde\psi_{\gamma}})$, and consider a mid-circuit measurement operation that measures the qubit $Q_i$ and writes the outcome value to the classical bit $C_j$. By isolating the qubit $Q_i$, the $n$-qubit state represented by $\ketfixed{\tilde\psi_{\gamma}}$ can be written as
\(\lambda_0\ketfixed{0}_{Q_i}\ketfixed{\phi_0}~+~\lambda_1\ketfixed{1}_{Q_i}\ketfixed{\phi_1},\)
where $\ketfixed{\phi_0}$ and $\ketfixed{\phi_1}$ are normalized states of the remaining $n-1$ qubits, and
$|\lambda_0|^2 + |\lambda_1|^2 = 1$. Measuring $Q_i$ in the computational basis, the outcome $m \in \{0, 1\}$ occurs with probability $p_m=|\lambda_m|^2$, and conditioned on the outcome $m$, the post-measurement state is $\ketfixed{m}_{Q_i}\ketfixed{\phi_m}$.

The quantum register state is updated as follow.
Let $P_0 = \ketfixed{0}\!\brafixed{0}$ and $P_1 = \ketfixed{1}\!\brafixed{1}$
denote the projectors onto the computational-basis states.
Let
\(P_m^{Q_i} = I^{\otimes(i-1)} \otimes P_m \otimes I^{\otimes(n-i)}\) be the corresponding projector acting on qubit $Q_i$ in the full $n$-qubit Hilbert space.
For calculating the probability $p_m$ we perform the tensor network contraction 
\(p_m =\brafixed{\tilde\psi_{\gamma}} P_m^{Q_i} \ketfixed{\tilde\psi_{\gamma}}\), 
where $\ketfixed{\tilde\psi_{\gamma}}$ is the tensor network representing the current quantum state, $\brafixed{\tilde\psi_{\gamma}}$ is its conjugate tensor network.
To apply the measurement operation to the qubit $Q_i$ we perform
\(\frac{1}{\sqrt{p_m}}P_m^{Q_i} \ketfixed{\tilde\psi_{\gamma}}\), with \(p_m>0\).
At the tensor-network level, the projector $P_m^{Q_i}$ is applied to the physical index associated with qubit \(Q_i\). This modifies only the local tensor carrying that physical leg, after which the state is renormalized by the factor \(1/\sqrt{p_m}\). Thus, the post-measurement update can be performed without reconstructing the full \(2^n\)-dimensional state vector.

On the other hand, the classical register is updated consistently with the observed outcome, by updating the state of the classical bit $C_j$ with the outcome value $m$. Consider the current classical state $c_{\gamma} = (c_1,\dots,c_m)$, then the updated classical state is
\((c_1,\dots,c_{j-1}, m, c_{j+1},\dots,c_m)\).
Thus, all classical bits except $C_j$ remain unchanged, and the value stored in $C_j$ is overwritten by the measurement outcome.

\subsubsection{Classically controlled operations}
\label{sec:method-dynamic-ops-ccop}
Consider an execution branch $(p_{\gamma}, c_{\gamma}, \ketfixed{\tilde\psi_{\gamma}})$, and a classically controlled operation consisting of a unitary $U$, conditioned on a Boolean expression over the bits of the classical register $C$.
Let $\varphi : \{0,1\}^m \to \{true, false\}$
be the Boolean function associated with the classical control expression. If $\varphi(c_{\gamma})=true$, the unitary $U$ is applied at runtime updating the quantum state as $U\ketfixed{\tilde\psi_{\gamma}}$, on the other hand, if $\varphi(c_{\gamma})=false$, the operation $U$ is not applied leaving the quantum state $\ketfixed{\tilde\psi_{\gamma}}$ unchanged. In our simulation process, whenever the condition is satisfied, the corresponding operation is treated exactly as a standard unitary gate and incorporated into the next DMRG compression step; otherwise, the operation is skipped. In this way, the simulation process captures the adaptive structure of dynamic circuits while preserving the same tensor-network compression scheme used in the purely unitary setting. Unlike mid-circuit measurements, classically controlled operations do not generate new execution branches. In the current branch, the operation is applied (or not) deterministically once the value of the classical condition has been evaluated.

\subsubsection{Reset operations}
\label{sec:method-dynamic-ops-res}
A reset is a unary non-unitary operation that forces the state of a qubit $Q_i$ to the basis state $\ketfixed{0}$.
Resetting a qubit $Q_i$ can be implemented as a measurement of $Q_i$ in the computational basis, followed by the conditional application of an $X$ gate to $Q_i$ whenever the measurement outcome is $1$. If the outcome is $0$, no further operation is applied. Therefore, after this procedure, the qubit is deterministically set in the state $\ketfixed{0}$. As a result, we can simulate reset operations using the mechanisms already introduced for mid-circuit measurements and classically controlled operations.

\subsection{Handling unitary chunks in dynamic circuits}
\label{sec:method-chunk-dyn-circ}
In \cref{subsec:prelim-sim-dmrg} we recalled that, for unitary circuits, the DMRG simulation proceeds through a sequence of compression steps, each associated with a unitary chunk of \(k\) consecutive gates. In this setting, the circuit is processed by repeatedly grouping k unitary operations into chunks and updating the current tensor-network state via a variational approximation. For dynamic circuits, this strategy must be adapted to account for mid-circuit measurements, which are non-unitary and cannot be absorbed into a standard DMRG compression step, and for classically controlled operations, whose Boolean conditions must be evaluated to determine whether the corresponding operations should be applied.

Let \(k\) denote the maximum size of a unitary chunk. A chunk is constructed by scanning the circuit operations sequentially. Unitary gates are always appended to the current chunk. For a classically controlled operation, the associated Boolean condition is evaluated on the current classical-register state: if it is satisfied, the corresponding unitary is appended; otherwise, the operation is skipped.
Chunk construction stops either when \(k\) operations have been collected or when a mid-circuit measurement is encountered. In the former case, the chunk is processed through a DMRG compression step. In the latter, the current chunk, if nonempty, is compressed through DMRG even if it contains fewer than \(k\) operations and the measurement is then applied to the current state before starting creating the new following chunk. In this way, we preserve the variational tensor-network mechanism used in the unitary case while extending it to the adaptive evolution induced by dynamic circuit components.

\subsection{DMRG simulation workflow for dynamic circuits}
\label{sec:method-dynamic-workflow}
As discussed in \cref{def:dynamic_circuit_state}, the state of a dynamic circuit is described by a collection of execution branches. Each branch is indexed by a measurement path \(\gamma\), namely the ordered sequence of outcomes produced by the mid-circuit measurements encountered along that branch. Because the number of active branches can grow exponentially with the number of mid-circuit measurements, their propagation requires dedicated simulation strategies. In this work, we consider two alternatives. A single-path strategy propagates only one branch at a time, thereby avoiding exponential memory growth but requiring repeated runs to recover ensemble statistics. A multi-path strategy instead propagates multiple active branches simultaneously, with the option of capping their number through probability-based truncation when needed.

\subsubsection{Single-path simulation}
\label{sec:single-path}
\begin{algorithm}[t]
\scriptsize
\SetInd{0.05em}{0.5em}
\caption{Single-path simulation \cref{sec:single-path}}
\label{alg:dynamic-single-path}
\KwIn{Dynamic circuit $D=(Q,C,O)$, maximum chunk size $k$}
\KwOut{Estimated fidelity $\tilde{\mathcal F}$ and final branch $(p_\gamma,c_\gamma,\ketfixed{\tilde\psi_\gamma})$}

$\ketfixed{\tilde\psi_\gamma} \leftarrow \ketfixed{0}^{\otimes n}$, $c_\gamma \leftarrow (0,\dots,0) \in \{0,1\}^m$, $p_\gamma \leftarrow 1$, \quad $\tilde{\mathcal F} \leftarrow 1$,
$\texttt{chunk} \leftarrow [\,]$\;

\ForEach{$O_\ell \in O$}{
    \If{$O_\ell$ is classically controlled and $\varphi(c_\gamma)=\mathrm{false}$}{
        \textbf{continue}\;
    }

    \If{$O_\ell$ is a mid-circuit measurement}{
        \If{$|\texttt{chunk}|>0$}{
            $(\ketfixed{\tilde\psi_\gamma}, f) \leftarrow \mathrm{Compress}(\ketfixed{\tilde\psi_\gamma}, \texttt{chunk})$\;
            $\tilde{\mathcal F} \leftarrow \tilde{\mathcal F}\, f$\;
            $\texttt{chunk} \leftarrow [\,]$\;
        }

        $(m,p_m,\ketfixed{\tilde\psi_\gamma}) \leftarrow \mathrm{Measure}(\ketfixed{\tilde\psi_\gamma}, O_\ell)$\;
        $p_\gamma \leftarrow p_\gamma\, p_m$\;
        $c_\gamma \leftarrow \mathrm{UpdateClassicalRegister}(c_\gamma, O_\ell, m)$\;
    }
    \Else{
        append $O_\ell$ to $\texttt{chunk}$\;
        \If{$|\texttt{chunk}|=k$}{
            $(\ketfixed{\tilde\psi_\gamma}, f) \leftarrow \mathrm{Compress}(\ketfixed{\tilde\psi_\gamma}, \texttt{chunk})$\;
            $\tilde{\mathcal F} \leftarrow \tilde{\mathcal F}\, f$\;
$\texttt{chunk} \leftarrow [\,]$\;
        }
    }
}

\If{$|\texttt{chunk}|>0$}{
    $(\ketfixed{\tilde\psi_\gamma}, f) \leftarrow \mathrm{Compress}(\ketfixed{\tilde\psi_\gamma}, \texttt{chunk})$\;
    $\tilde{\mathcal F} \leftarrow \tilde{\mathcal F}\, f$\;
}

\Return $(\tilde{\mathcal F},p_\gamma,c_\gamma,\ketfixed{\tilde\psi_\gamma})$\;
\end{algorithm}
The single-path approach provides a stochastic simulation strategy for dynamic circuits. Each run retains only one measurement path \(\gamma\), obtained by sampling mid-circuit measurement outcomes according to their probabilities.
The procedure is given in \cref{alg:dynamic-single-path}.
The algorithm takes as input a dynamic circuit \(D=(Q,C,O)\) and a parameter \(k\), interpreted as the maximum size of a unitary chunk. It returns the tuple
\((\tilde{\mathcal F}, p_\gamma, c_\gamma, \ketfixed{\tilde\psi_\gamma}),\)
where \(p_\gamma\) is the probability of the sampled path \(\gamma\), \(c_\gamma\) is the corresponding final classical-register configuration, \(\ketfixed{\tilde\psi_\gamma}\) is the tensor-network approximation of the final quantum state, and \(\tilde{\mathcal F}\) is the estimated fidelity accumulated along the simulation.
The procedure is initialized with \(\ketfixed{\tilde\psi_\gamma}\) equal to a tensor-network representation of the all-zero quantum state \(\ketfixed{0}^{\otimes n}\), with the classical register set to \(c_\gamma=(0,\dots,0)\), path probability \(p_\gamma=1\), estimated fidelity \(\tilde{\mathcal F}=1\), and with an empty unitary chunk \(\texttt{chunk}=[\,]\). The operations in \(O\) are then processed sequentially.

Unitary operations are not applied individually. Instead, they are appended to the current chunk. Whenever the chunk reaches size \(k\), the routine
\[
(\ketfixed{\tilde\psi_\gamma}, f)\leftarrow \mathrm{Compress}(\ketfixed{\tilde\psi_\gamma},\texttt{chunk})
\]
is invoked. As discussed in \cref{subsec:prelim-sim-dmrg}, this routine applies the unitary chunk to the current tensor-network state, variationally compresses the result within the chosen ansatz, and returns both the updated state and the corresponding partial fidelity \(f\). The estimated fidelity is then updated as
\(
\tilde{\mathcal F}\leftarrow \tilde{\mathcal F}\,f,
\)
and the chunk is set again to empty.

Classically controlled operations are handled using the current classical-register state, as described in \cref{sec:method-dynamic-ops-ccop}. If the associated Boolean condition evaluates to false on \(c_\gamma\), the operation is skipped. Otherwise, it is treated as an ordinary unitary operation and appended to the current chunk.

When a mid-circuit measurement is encountered, the current chunk is first compressed if it is nonempty. Suppose that the measurement acts on qubit \(Q_i\) and writes the observed outcome to the classical bit \(C_j\). As discussed in \cref{sec:method-dynamic-ops-mcm}, the measurement is processed through
\[
(m,p_m,\ketfixed{\tilde\psi_\gamma})\leftarrow \mathrm{Measure}(\ketfixed{\tilde\psi_\gamma},O_\ell),
\]
where \(m\in\{0,1\}\) is the sampled outcome,
\(p_m=\brafixed{\tilde\psi_\gamma} P_m^{Q_i}\ketfixed{\tilde\psi_\gamma}\)
is its probability, and
\(
\ketfixed{\tilde\psi_\gamma}\leftarrow \frac{P_m^{Q_i}\ketfixed{\tilde\psi_\gamma}}{\sqrt{p_m}}
\)
is the corresponding post-measurement state. The classical register is then updated by writing the sampled outcome \(m\) into the \(j\)-th bit, i.e.,
\(
c_\gamma=(c_1,\dots,c_{j-1},m,c_{j+1},\dots,c_m),
\)
equivalently implemented in the pseudocode through
\[
c_\gamma\leftarrow \mathrm{UpdateClassicalRegister}(c_\gamma,O_\ell,m).
\]
Finally, the path probability is updated as \(p_\gamma\leftarrow p_\gamma\,p_m\).

After all operations have been processed, any residual nonempty chunk is compressed and its partial fidelity is incorporated into \(\tilde{\mathcal F}\). The final output is therefore the simulated branch associated with the sampled measurement path.


\subsubsection{Multi-path simulation}
\label{sec:multiple-path}
\begin{algorithm}[t]
\scriptsize
\SetInd{0.05em}{0.5em}
\caption{Multi-path simulation (\cref{sec:multiple-path})}
\label{alg:dynamic-multi-path}
\KwIn{Dynamic circuit $D=(Q,C,O)$, maximum chunk size $k$, maximum number of retained paths $M$}
\KwOut{Final list of active branches $\mathcal{L}$}

$\mathcal{L} \leftarrow [(\gamma=\emptyset,\; p_\gamma\leftarrow1,\; c_\gamma\leftarrow(0,\dots,0),\; \ketfixed{\tilde\psi_\gamma}\leftarrow\ketfixed{0}^{\otimes n},\; \tilde{\mathcal F}_\gamma\leftarrow 1,\; \texttt{chunk}_\gamma\leftarrow[\,])]$\;

\ForEach{$O_\ell \in O$}{
    $\mathcal{L}_{\mathrm{next}} \leftarrow [\,]$\;

    \ForEach{$( \gamma,p_\gamma,c_\gamma,\ketfixed{\tilde\psi_\gamma},\tilde{\mathcal F}_\gamma,\texttt{chunk}_\gamma ) \in \mathcal{L}$}{
        \If{$O_\ell$ is classically controlled and $\varphi(c_\gamma)=\mathrm{false}$}{
            append $( \gamma,p_\gamma,c_\gamma,\ketfixed{\tilde\psi_\gamma},\tilde{\mathcal F}_\gamma,\texttt{chunk}_\gamma )$ to $\mathcal{L}_{\mathrm{next}}$\;
            \textbf{continue}\;
        }

        \If{$O_\ell$ is a mid-circuit measurement}{
            \If{$|\texttt{chunk}_\gamma|>0$}{
                $(\ketfixed{\tilde\psi_\gamma}, f) \leftarrow \mathrm{Compress}(\ketfixed{\tilde\psi_\gamma}, \texttt{chunk}_\gamma)$\;
                $\tilde{\mathcal F}_\gamma \leftarrow \tilde{\mathcal F}_\gamma\, f$\;
                $\texttt{chunk}_\gamma \leftarrow [\,]$\;
            }

            \ForEach{$m \in \{0,1\}$ with $p_m \neq 0$}{
                $(\ketfixed{\tilde\psi_{\gamma m}}, p_m) \leftarrow \mathrm{Collapse}(\ketfixed{\tilde\psi_\gamma}, O_\ell, m)$\;
                $p_{\gamma m} \leftarrow p_\gamma\, p_m$\;
                $c_{\gamma m} \leftarrow \mathrm{UpdateClassicalRegister}(c_\gamma, O_\ell, m)$\;
                $\tilde{\mathcal F}_{\gamma m} \leftarrow \tilde{\mathcal F}_\gamma$\;
                $\texttt{chunk}_{\gamma m} \leftarrow [\,]$\;
                append $(\gamma m,p_{\gamma m},c_{\gamma m},\ketfixed{\tilde\psi_{\gamma m}},\tilde{\mathcal F}_{\gamma m},\texttt{chunk}_{\gamma m})$ to $\mathcal{L}_{\mathrm{next}}$\;
            }
        }
        \Else{
            append $O_\ell$ to $\texttt{chunk}_\gamma$\;
            \If{$|\texttt{chunk}_\gamma|=k$}{
                $(\ketfixed{\tilde\psi_\gamma}, f) \leftarrow \mathrm{Compress}(\ketfixed{\tilde\psi_\gamma}, \texttt{chunk}_\gamma)$\;
                $\tilde{\mathcal F}_\gamma \leftarrow \tilde{\mathcal F}_\gamma\, f$\;
                $\texttt{chunk}_\gamma \leftarrow [\,]$\;
            }
            append $( \gamma,p_\gamma,c_\gamma,\ketfixed{\tilde\psi_\gamma},\tilde{\mathcal F}_\gamma,\texttt{chunk}_\gamma )$ to $\mathcal{L}_{\mathrm{next}}$\;
        }
    }

    $\mathcal{L} \leftarrow \mathcal{L}_{\mathrm{next}}$\;

    \If{$O_\ell$ is a mid-circuit measurement and $|\mathcal{L}|>M$}{
        retain in $\mathcal{L}$ only the $M$ branches with largest values of $p_\gamma$\;
    }
}

\ForEach{$( \gamma,p_\gamma,c_\gamma,\ketfixed{\tilde\psi_\gamma},\tilde{\mathcal F}_\gamma,\texttt{chunk}_\gamma ) \in \mathcal{L}$}{
    \If{$|\texttt{chunk}_\gamma|>0$}{
        $(\ketfixed{\tilde\psi_\gamma}, f) \leftarrow \mathrm{Compress}(\ketfixed{\tilde\psi_\gamma}, \texttt{chunk}_\gamma)$\;
        $\tilde{\mathcal F}_\gamma \leftarrow \tilde{\mathcal F}_\gamma\, f$\;
        $\texttt{chunk}_\gamma \leftarrow [\,]$\;
    }
}

\Return $\mathcal{L}$\;
\end{algorithm}

The multi-path approach propagates multiple measurement paths simultaneously. Instead of retaining a single sampled path, it maintains a list of active branches and updates them in parallel as the circuit is scanned.
The complete procedure is given in \cref{alg:dynamic-multi-path}.
The algorithm takes as input a dynamic circuit \(D=(Q,C,O)\), a parameter \(k\) specifying the maximum size of a unitary chunk, and a parameter \(M\) specifying the maximum number of retained paths. Its output is a list
\(
\mathcal{L}=\{(\gamma,p_\gamma,c_\gamma,\ketfixed{\tilde\psi_\gamma},\tilde{\mathcal F}_\gamma)\}_\gamma
\)
of active branches, where each branch is indexed by a  path \(\gamma\) and carries its corresponding path probability, classical-register configuration, tensor-network state, and estimated fidelity.
The procedure is initialized with a single branch with quantum state \(\ketfixed{0}^{\otimes n}\), classical register \(c_\gamma=(0,\dots,0)\), path probability \(p_\gamma=1\), estimated fidelity \(\tilde{\mathcal F}_\gamma=1\), and empty chunk \(\texttt{chunk}_\gamma=[\,]\). At each step, the list of active branches \(\mathcal{L}\) is updated into a new list \(\mathcal{L}_{\mathrm{next}}\).

For each active branch, unitary operations and classically controlled operations are handled exactly as in the single-path setting. In particular, applicable unitary operations are appended to the branch-local chunk \(\texttt{chunk}_\gamma\), and whenever the chunk reaches size \(k\), the routine
\[
(\ketfixed{\tilde\psi_\gamma},f)\leftarrow \mathrm{Compress}(\ketfixed{\tilde\psi_\gamma},\texttt{chunk}_\gamma)
\]
is applied to that branch. The corresponding estimated fidelity is then updated as \(\tilde{\mathcal F}_\gamma\leftarrow \tilde{\mathcal F}_\gamma\,f\), and the chunk is reset. Classically controlled operation conditions are evaluated on the branch classical-register state $c_{\gamma}$.

The main difference with respect to the single-path approach arises at mid-circuit measurements.
Instead of sampling one outcome, the algorithm generates all outcomes \(m\in\{0,1\}\) with nonzero probability. For each such outcome, the routine
\[
(\ketfixed{\tilde\psi_{\gamma m}},p_m)\leftarrow \mathrm{Collapse}(\ketfixed{\tilde\psi_\gamma},O_\ell,m)
\]
returns the post-measurement state associated with outcome \(m\) together with its conditional probability \(p_m\), computed as described in \cref{sec:method-dynamic-ops-mcm}. A new branch indexed by the extended path \(\gamma m\), for each $m \in \{0,1\}$, is then created, with updated path probability \(p_{\gamma m}=p_\gamma p_m\), updated quantum register $\ketfixed{\tilde\psi_{\gamma m}}$, updated classical register \(c_{\gamma m}\), and estimated fidelity \(\tilde{\mathcal F}_{\gamma m}=\tilde{\mathcal F}_\gamma\).
Since the number of active paths can grow exponentially with the number of measurements, the algorithm optionally truncates the branch list after each measurement step. More precisely, if the number of active branches exceeds the threshold \(M\), only the \(M\) branches with largest values of \(p_\gamma\) are retained. This truncation introduces an additional approximation, beyond tensor-network compression, by discarding low-probability paths from the simulated ensemble. 

After all operations have been processed, any residual nonempty chunk is compressed independently on each retained branch. The final output is therefore the list of retained branches produced by the multi-path evolution, together with their associated fidelity.
\subsection{Correctness and complexity analysis}
\label{sec:method-correctness-complexity}

\subsubsection{Correctness}
Both simulation strategies are consistent with the branch-based semantics of dynamic circuits (\cref{def:dynamic_circuit_state}). In both cases, unitary operations are processed through the same DMRG compression steps used in the unitary-circuit setting \cite{ayral2023simulationmps, dubey2025simulating}, while classically controlled operations are evaluated on the current classical-register state and either applied or skipped accordingly. Mid-circuit measurements are handled by the measurement-update rule (\cref{sec:method-dynamic-ops-mcm}), which updates both the quantum state and the classical register.
In the single-path strategy, each run propagates one valid execution branch obtained by sampling measurement outcomes according to their probabilities, and repeated runs sample measurement paths according to the probability distribution induced by the dynamic circuit semantics. In the multi-path strategy, each mid-circuit measurement explicitly generates all outcomes, so that the algorithm reconstructs the full collection of reachable branches. The only sources of approximation in our framework are the tensor-network compressions performed for each unitary chunk, controlled by the chosen bond dimensions, and in the multi-path setting, the optional truncation of low-probability branches.

\subsubsection{Complexity}
At the branch level, the dominant cost of both strategies is the DMRG-based simulation of unitary chunks, whose complexity depends on the chosen tensor-network ansatz and on the corresponding bond dimensions, as in the unitary-circuit setting \cite{ayral2023simulationmps, dubey2025simulating}. The additional cost due to dynamic operations is comparatively modest: classically controlled operations only require evaluating a Boolean condition, while a measurement requires a tensor-network contraction and a local state update, which is typically less expensive than the sequence of contractions involved in a compression step.
The single-path strategy stores and updates only one active branch, so its memory cost is that of a single tensor-network state and its runtime is linear in the circuit length for each run, up to the cost of compression and measurement updates. In contrast, the multi-path strategy processes all active branches simultaneously. If the circuit contains \(r\) mid-circuit measurements, the number of branches can grow up to \(2^r\) in the absence of truncation, leading to exponential time and memory growth. When truncation is enabled, the number of retained branches is bounded by \(M\), so both runtime and memory scale linearly in \(M\).

\section{Evaluation}
\label{sec:result}
In this section, we experimentally evaluate the proposed simulator on quantum teleportation (\cref{subsec:evaluation_teleportation}), GHZ state preparation (\cref{sec:evaluation_ghz}), and random dynamic circuits (\cref{sec:evaluation_random})\footnote{Code available on Github: \url{https://github.com/1nnocenzo/tn-dc-sim}}. The experiments are performed on a machine equipped with an Apple M1 Pro chip and 16GB of RAM.
\subsection{Quantum teleportation protocol}
\label{subsec:evaluation_teleportation}
\begin{figure}[t]
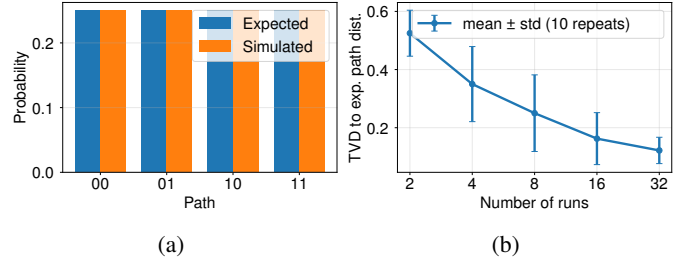

  \centering

  \subfloat[]{%
    \resizebox{0.5\linewidth}{!}{\input{images/validation_all_branches_teleportation_probs.pgf}}%
    \label{subfig:telep-all-no-bob-meas}%
  }\hfill
  \subfloat[]{%
    \resizebox{0.5\linewidth}{!}{\input{images/validation_single_branch_teleportation_tvd_vs_n.pgf}}%
    \label{subfig:telep-single-no-bob-meas}%
  }
  \caption{Teleportation protocol benchmark without final measurement on Bob's qubit.
\cref{subfig:telep-all-no-bob-meas} reports the expected and simulated probabilities of the four execution paths using the multi-path simulator.
\cref{subfig:telep-single-no-bob-meas} reports the TVD between the expected and the empirical path distribution obtained with the single-path simulator for different numbers of runs.}
  \label{fig:teleport_no_final_meas}
\end{figure}

We first evaluate our simulation workflow on the quantum teleportation protocol \cite{teleportation_Bennett_1895}, using the MPS representation and fixed simulator parameters (chunk size $k=20$, number of sweeps $s=2$, and bond dimension $\chi=2$) that allow us to correctly simulate the expected circuit semantics. With this benchmark we validate the correctness of the proposed branch-based simulation. In the teleportation protocol, Alice teleports an unknown single-qubit state to Bob by consuming a shared Bell pair, measuring her two qubits, and communicating the two classical outcomes to Bob, who applies the corresponding classically controlled correction gates to reconstruct the input state on his qubit. The outcomes of the two mid-circuit measurements define four possible execution paths 
\(
\gamma \in \{00,01,10,11\}.
\)
that occur with equal probability
\(p_\gamma = 0.25\). The subsequent classically controlled operations are then conditioned on the observed path.
We consider a circuit that teleports the qubit defined as
\(
\ket{\psi}
=
0.61\ket{0}+(0.59 + 0.53\,i)\ket{1}.
\)
We verify that the multi-path simulator explicitly reconstructs all four execution branches with the expected probabilities. As shown in \cref{subfig:telep-all-no-bob-meas}, the probabilities assigned by the simulator match the ideal branch distribution of the teleportation circuit. This confirms that the multi-path simulation method correctly generates the four paths together with their corresponding probabilities.
We then evaluate the single-path simulator, which samples one measurement path per execution and therefore requires multiple independent runs to estimate the full branch distribution. Given \(N\) independent single-path runs, we denote by \(N_\gamma\) the number of runs in which the measurement path \(\gamma\) is sampled. The empirical probability assigned to path \(\gamma\) is therefore
\(\hat p_\gamma = \frac{N_\gamma}{N}.\)
We compare the empirical distribution \(\hat p\) obtained from repeated single-path runs with the ideal distribution \(p\) over the four paths using the total variation distance 
\(
\mathrm{TVD}(p,\hat p)
=
\frac{1}{2}
\sum_{\gamma}
\left|p_\gamma - \hat p_\gamma\right|
\)\cite{LevinPeresWilmer2006}.
We vary the number of single-path runs \(N\) from \(1\) to \(32\), using powers of 2, and repeat each configuration \(10\) times. The results are reported as mean and standard deviation across these repetitions. As shown in \cref{subfig:telep-single-no-bob-meas}, the $\mathrm{TVD}$ decreases as \(N\) increases, indicating that the empirical distribution sampled by the single-path simulator converges toward the ideal teleportation branch distribution.

\begin{figure}[t]
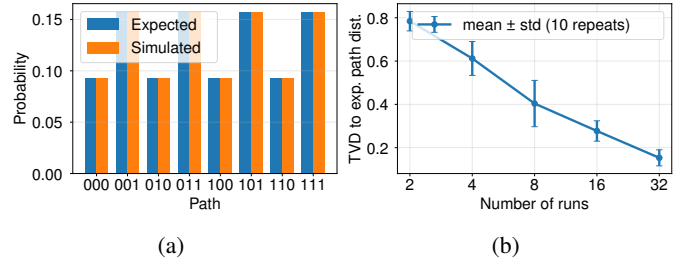

  \centering

  \subfloat[]{%
    \resizebox{0.5\linewidth}{!}{\input{images/validation_all_branches_teleportation_bob_measure_probs.pgf}}%
    \label{subfig:telep-all-bob-meas}%
  }\hfill
  \subfloat[]{%
    \resizebox{0.5\linewidth}{!}{\input{images/validation_single_branch_teleportation_bob_measure_tvd_vs_n.pgf}}%
    \label{subfig:telep-single-bob-meas}%
  }
  \caption{Teleportation protocol benchmark with final measurement on Bob's qubit.
  \cref{subfig:telep-all-bob-meas} reports the expected and simulated probabilities of the eight execution paths using the multi-path simulator.
  \cref{subfig:telep-single-bob-meas} reports the TVD between the expected and the empirical path distribution obtained with the single-path simulator for different numbers of runs.}
  \label{fig:telep-single-bob-meas}
\end{figure}

To verify that the quantum states associated with the different execution branches are propagated correctly, we also consider a variant of the experiment in which a final measurement is applied to Bob's qubit after the classically controlled corrections. Since the teleported state is $0.61\ket{0}+(0.59 + 0.53\,i)\ket{1}$
the expected probabilities of measuring Bob's qubit in the computational basis are
\(\Pr(0)=|\alpha|^2 \approx 0.37,\,
\Pr(1)=|\beta|^2 \approx 0.63.\)
Combining this final measurement with the four possible mid-circuit measurement paths produces eight final execution paths: four paths ending with Bob's qubit measured as \(0\), each with probability approximately \(0.25\cdot 0.37 = 0.0925\), and four paths ending with Bob's qubit measured as \(1\), each with probability approximately \(0.25\cdot 0.63 = 0.1575\). As shown in \cref{subfig:telep-all-bob-meas}, the multi-path simulator reproduces this expected distribution over the eight final paths, showing that the simulator also preserves the quantum state associated with each branch up to the final measurement.
We repeat the experiment with the single-path simulator, estimating the empirical distribution from multiple independent runs. As shown in \cref{subfig:telep-single-bob-meas}, the TVD between the empirical distribution and the expected eight-path distribution decreases as the number of runs increases, confirming statistical convergence.




\subsection{Dynamic procedure for GHZ state preparation}
\label{sec:evaluation_ghz}
\begin{figure}[t]
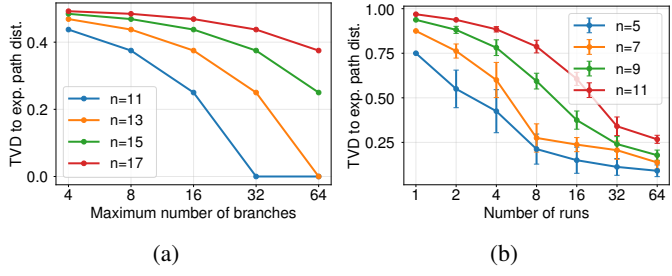

  \centering
  \subfloat[]{%
    \resizebox{0.495\linewidth}{!}{\input{images/ghz_pruning_tvd_vs_maxbranches.pgf}}%
    \label{subfig:ghz-max_branches_tvd}%
  }
  \hfill
  \subfloat[]{%
    \resizebox{0.495\linewidth}{!}{\input{images/ghz_single_vs_all_tvd_vs_nshots.pgf}}%
    \label{subfig:ghz_single_vs_all_tvd_vs_nshots}%
  }
  \caption{TVD between the simulated and expected path distributions for the dynamic GHZ preparation benchmark. 
\cref{subfig:ghz-max_branches_tvd} shows the multi-path simulation with branch truncation. 
 \cref{subfig:ghz_single_vs_all_tvd_vs_nshots} shows the single-path simulation, reporting the mean and standard deviation of $10$ executions for each settings. }
  \label{fig:ghz-branch-distr}
\end{figure}

As a second benchmark, we evaluate the simulator on a dynamic procedure for preparing \(n\)-qubit GHZ states
that follows a constant-depth construction in which entanglement is first generated locally and then extended through mid-circuit measurements and classically controlled operations \cite{baumer2024efficient}.
For an input size \(n\), with \(n\) odd, the circuit contains
\(r = \frac{n-1}{2}\)
mid-circuit measurements, one for each even-indexed qubit, namely
\(Q_2,Q_4,\dots,Q_{n-1}\). The same number of reset operations is applied after the feed-forward stage. However, these reset operations do not introduce additional paths, since they are applied to qubits that have already been measured in the computational basis. Their effect is therefore deterministic within each branch. The circuit also contains
$r$
classically controlled \(X\) corrections. Since each mid-circuit measurement can produce two possible outcomes, the dynamic execution can generate up to
\(2^r = 2^{(n-1)/2}\)
distinct execution paths with equal probability.
In our experiments, we instantiate the dynamic GHZ preparation circuit for
\(n \in \{5,7,9,11,13,15,17\}\),
corresponding to \(r \in \{2,3,4,5,6,7,8\}\) mid-circuit measurements and \(2^r\) equiprobable execution paths, with expected probability \(p_\gamma=2^{-r}\) for each path \(\gamma\).

We evaluate both simulation strategies using an MPS representation with fixed simulator parameters ($k=25$, $s=2$, and $\chi=4$), as in the teleportation benchmark, by comparing the simulator’s path distribution with the ideal distribution of the dynamic GHZ preparation circuit.
\cref{subfig:ghz-max_branches_tvd} reports the results for the multi-path simulator. In this case, the approximation is controlled by the maximum number \(M\) of retained branches. When \(M\) is smaller than the total number of reachable paths, the simulator keeps only a subset of the full branch ensemble, and the TVD reflects the probability mass associated with the discarded paths. Accordingly, the distance decreases as \(M\) increases. For smaller instances, such as \(n=11\) and \(n=13\), the TVD reaches zero once the branch cap is large enough to retain all reachable paths. For larger instances, such as \(n=15\) and \(n=17\), the number of possible paths exceeds the largest tested value of \(M\), and the TVD remains strictly positive even for \(M=64\).
\cref{subfig:ghz_single_vs_all_tvd_vs_nshots} shows the corresponding results for the single-path simulator. Here, each execution samples only one measurement path, and the empirical distribution is estimated from repeated independent runs. As the number of runs increases, the TVD decreases for all system sizes, showing convergence toward the expected path distribution. However, the convergence becomes slower as \(n\) increases, since the number of possible paths grows exponentially as \(2^r\). Consequently, for larger circuits the same number of runs covers only a smaller fraction of the path space, leading to higher TVD values.
\begin{figure}[t]
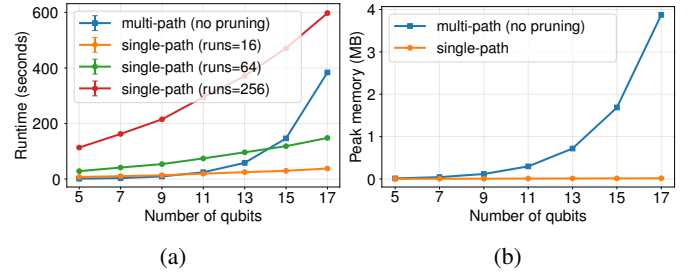

  \centering

  \subfloat[]{%
    \resizebox{0.5\linewidth}{!}{\input{images/ghz_runtime_vs_qubits_single_multi_10attempts.pgf}}%
    \label{subfig:ghz-runtime}%
  }\hfill
  \subfloat[]{%
    \resizebox{0.5\linewidth}{!}{\input{images/ghz_peak_memory_vs_qubits.pgf}}%
    \label{subfig:ghz-memory}%
  }
  \caption{Computational cost of the dynamic GHZ preparation benchmark: runtime (\cref{subfig:ghz-runtime}) and peak memory usage (\cref{subfig:ghz-memory}) as a function of qubit count, averaged over 10 independent runs.}
  \label{fig:ghz_runtime_memory}
\end{figure}
We also compare the computational cost of the two simulation strategies on the dynamic GHZ benchmark.
The runtime results are shown in \cref{subfig:ghz-runtime}. In the multi-path setting, we disable branch pruning, so that the simulator eventually propagates the complete set of \(2^r\) execution paths. The runtime therefore grows  with the number of qubits, reflecting the exponential increase in the final number of branches. However, the cost of multi-path simulation is not equivalent to running one independent simulation for each final path. Branches are created only when mid-circuit measurements are encountered: before the first measurement the simulator propagates a single state, and the number of active branches increases progressively as the measurements are processed. As a result, circuit portions that are common to all execution paths are evaluated only once. In contrast, repeated single-path simulation executes the entire circuit independently for each run, recomputing also the common prefix before any branching occurs. This explains why multi-path simulation remains competitive for moderate system sizes, even without pruning. The single-path runtime, instead, is controlled by the number of independent runs and grows accordingly when increasing the number of sampled executions from \(16\) to \(64\) and \(256\).
\cref{subfig:ghz-memory} reports the corresponding peak memory usage. The multi-path simulator shows an increasing memory usage as \(n\) grows, due to the need to store multiple active tensor-network states, one for each branch. In contrast, the single-path simulator stores only one active branch at a time, and its peak memory remains essentially constant across the tested system sizes.

\begin{figure*}
    \centering
    \begin{subfigure}{.4\textwidth}
      \centering
      \includegraphics[width=\linewidth]{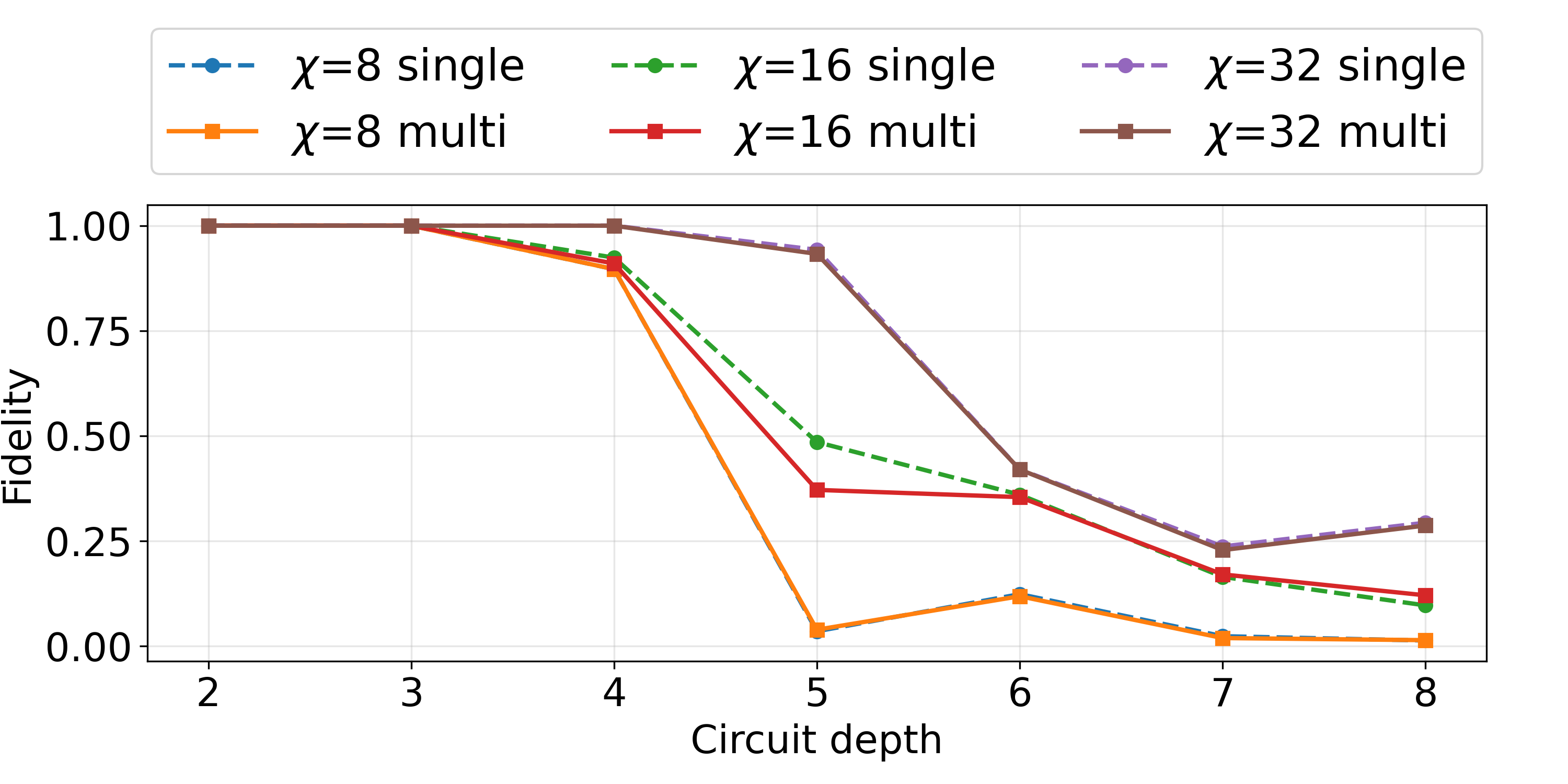}
      \caption{}
      \label{fig:mps_fidelity_vs_depth}
    \end{subfigure}%
    \begin{subfigure}{.4\textwidth}
      \centering
      \includegraphics[width=\linewidth]{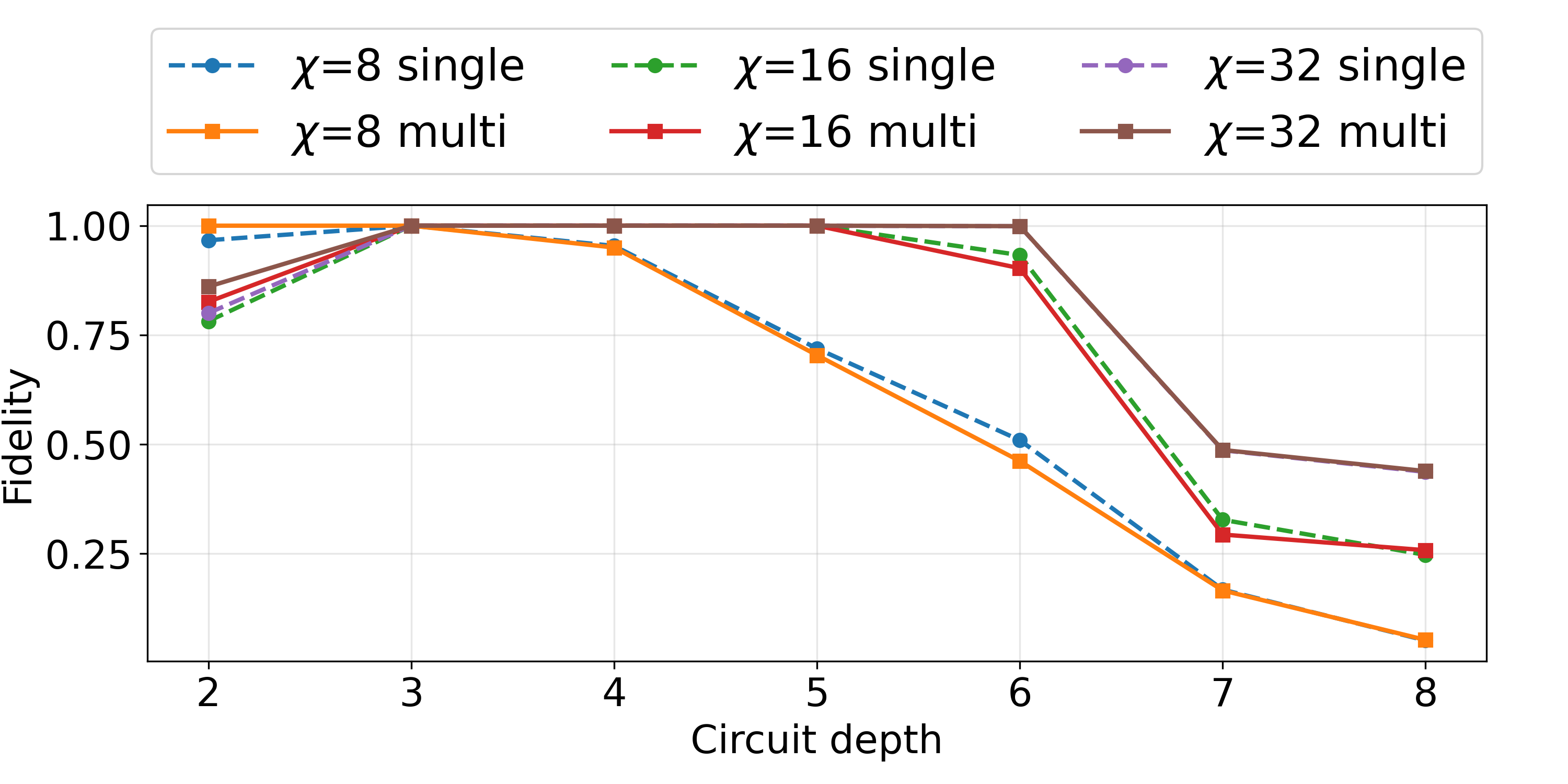}
      \caption{}
      \label{fig:ttn_fidelity_vs_depth}
    \end{subfigure}
    \caption{Simulation fidelity for 27-qubit random dynamic circuits (see \cref{tab:random_circuits}), using MPS (a) and TTN (b) representations.}
    \label{fig:fidelity_vs_depth_random}
\end{figure*}

\begin{table}[]
\centering
\scriptsize
\begin{tabular}{cccc}
\hline
depth & size & measurements op. & conditional op. \\
\hline
2 & 112.33 $\pm$ 7.23 & 11.33 $\pm$ 9.74 & 2.67 $\pm$ 1.25 \\
3 & 154.67 $\pm$ 18.96 & 5.00 $\pm$ 2.16 & 5.00 $\pm$ 3.56 \\
4 & 204.00 $\pm$ 11.00 & 15.00 $\pm$ 11.05 & 14.00 $\pm$ 9.27 \\
5 & 268.00 $\pm$ 11.18 & 18.33 $\pm$ 7.59 & 15.67 $\pm$ 8.06 \\
6 & 333.33 $\pm$ 12.47 & 18.33 $\pm$ 7.59 & 15.67 $\pm$ 8.06 \\
7 & 383.67 $\pm$ 7.41 & 18.67 $\pm$ 7.32 & 16.00 $\pm$ 7.87 \\
8 & 405.67 $\pm$ 26.44 & 24.67 $\pm$ 8.34 & 15.33 $\pm$ 6.94 \\
\hline
\end{tabular}
\caption{Properties of the random dynamic circuits used for evaluation; 3 circuits are generated for each depth.}
\label{tab:random_circuits}
\end{table}

\begin{table*}[t!]
\centering
\scriptsize
\begin{tabular}{c|llll|llll}
\hline
 & \multicolumn{4}{c|}{MPS} & \multicolumn{4}{c}{TTN} \\
\hline
depth & runtime sp [s] & memory sp [MB] & runtime mp [s] & memory mp [MB] & runtime sp [s] & memory sp [MB] & runtime mp [s] & memory mp [MB] \\
\hline
2 & 203.0 $\pm$ 22.2 & 1.20 $\pm$ 0.00 & 72.10 $\pm$ 2.49 & 19.22 $\pm$ 0.00 & 89.6 $\pm$ 2.1 & 2.64 $\pm$ 0.00 & 30.74 $\pm$ 0.20 & 42.24 $\pm$ 0.00 \\
3 & 296.5 $\pm$ 0.7 & 1.20 $\pm$ 0.00 & 57.31 $\pm$ 2.75 & 9.61 $\pm$ 0.00 & 125.4 $\pm$ 1.6 & 2.64 $\pm$ 0.00 & 23.02 $\pm$ 0.08 & 21.12 $\pm$ 0.00 \\
4 & 277.6 $\pm$ 11.1 & 1.20 $\pm$ 0.00 & 34.10 $\pm$ 0.52 & 4.81 $\pm$ 0.00 & 133.8 $\pm$ 1.4 & 2.64 $\pm$ 0.00 & 16.49 $\pm$ 1.06 & 10.56 $\pm$ 0.00 \\
5 & 366.9 $\pm$ 4.3 & 1.20 $\pm$ 0.00 & 104.63 $\pm$ 0.08 & 19.22 $\pm$ 0.00 & 174.5 $\pm$ 3.7 & 2.64 $\pm$ 0.00 & 46.65 $\pm$ 0.18 & 42.24 $\pm$ 0.00 \\
6 & 503.6 $\pm$ 3.1 & 1.20 $\pm$ 0.00 & 197.88 $\pm$ 4.24 & 19.22 $\pm$ 0.00 & 227.8 $\pm$ 1.9 & 2.64 $\pm$ 0.00 & 84.71 $\pm$ 2.62 & 42.24 $\pm$ 0.00 \\
7 & 565.2 $\pm$ 15.2 & 1.20 $\pm$ 0.00 & 233.97 $\pm$ 5.41 & 19.22 $\pm$ 0.00 & 252.0 $\pm$ 1.4 & 2.64 $\pm$ 0.00 & 99.10 $\pm$ 0.69 & 42.24 $\pm$ 0.00 \\
8 & 573.3 $\pm$ 29.7 & 1.20 $\pm$ 0.00 & 203.30 $\pm$ 86.80 & 16.02 $\pm$ 4.53 & 262.5 $\pm$ 24.5 & 2.64 $\pm$ 0.00 & 90.49 $\pm$ 38.52 & 35.20 $\pm$ 9.96 \\
\hline
\end{tabular}
\caption{Simulation costs using single-path (sp) and multi-path (mp) methods on random dynamic circuits with $\chi=32$.}
\label{tab:dynamic-scaling-combined-dmax-32}
\end{table*}

\subsection{Random dynamic circuits}
\label{sec:evaluation_random}

We evaluate our method using random dynamic circuit generated with an extended Qiskit's random circuit generator \cite{ibmRandom}.
Compared to the original version where dynamic components consist of measuring all the qubits and adding one classically controlled operation, our circuit generator yields more different patterns: it measures subsets of qubits of varying size and insert one or more classically controlled operations. For these experiments, we use both MPS and TTN state representation and we analyze the trade-off in estimated fidelity against computational cost, memory usage, and bond-dimension approximations, for different circuit depths. We set the number of qubits to 27, fixed compression settings (chunk size $k=20$, number of sweeps $s=2$), repeating the single-path method 10 times, and limiting the multi-path method to $M=8$ branches. Every run is repeated 3 times with different initialization (random circuit) (see \cref{tab:random_circuits}), while the tree structure for TTN is $[1,3,9,27]$. Figure \ref{fig:fidelity_vs_depth_random} reports fidelity as a function of circuit depth for $\chi\in\{8,16,32\}$, for both MPS (Fig.~\ref{fig:mps_fidelity_vs_depth}) and TTN (Fig.~\ref{fig:ttn_fidelity_vs_depth}).  
As expected, fidelity decreases with depth for all methods, while increasing $\chi$ significantly delays this degradation, especially in the deep-circuit regime. Note that here we are interested in the fidelity of branches propagation during the simulation, not in the path probability distribution as in \cref{fig:ghz-branch-distr}. While the reported fidelity for single path is the one returned by \cref{alg:dynamic-single-path}, for multi-path we normalize the final branch fidelities by the branch probabilities. For this reason we observe that single-path and multi-path curves behave similarly, meaning that the two approaches are coherent. Moreover, despite using a small value for $s$, the algorithms tend to return satisfactory fidelities, especially for low depth and high $\chi$, while increasing this parameter could improve the results for difficult configurations. In Table \ref{tab:dynamic-scaling-combined-dmax-32}, we report runtime and memory versus depth (and circuit size) at $\chi=32$ for MPS and TTN, respectively. Per run, multi-path is consistently slower and more memory demanding than single-path since multiple branches are propagated simultaneously. However, single-path runtimes are reported as totals over 10 repetitions; under this metric, they are typically higher than a single multi-path run. We use this aggregation for a fair comparison, noting that single-path iterations are trivially parallelizable. The single-path peak memory is  constant with depth because, for fixed network structure and $\chi$, tensor dimensions are bounded and only one branch is kept in memory at a time. By contrast, multi-path memory grows with depth as several branches coexist (up to the branch cap $M$).

Comparing tensor-network structures, we observe a clear runtime--memory trade-off: TTN is faster than MPS (both in single- and multi-path), while MPS is more memory efficient.  
Overall, these results suggest that $\chi$ is the primary control parameter for fidelity, while the choice between single- or multi-path approaches and MPS or TTN formats depends on whether the priority is throughput, memory efficiency, or fully branch-resolved dynamics.

\section{Conclusion}
\label{sec:conclusion}
In this work we provided a method for simulating dynamic circuits using a tensor network approach based on DMRG. We proposed two simulation algorithms: a single-path strategy, which for each measurement stochastically samples one outcome per run and propagates a single execution branch, and a multi-path strategy, which maintains an ensemble of branches simultaneously and optionally prunes low-probability paths when their number exceeds a given threshold. Validating our proposal using both standard dynamic circuit protocols like quantum teleportation and GHZ state preparation, and random circuits, we observed that both strategies correctly reproduce the quantum-state semantics of dynamic circuits. Results show the trade-off between the two strategies: single-path offers low per-run cost, while multi-path provides deterministic branch coverage at higher memory and runtime cost. 

While our method is designed to support MPS and TTN topologies, extending it to other tensor network architectures is a natural direction for further development. Other possible future works include further investigation into adaptive branch pruning strategies, beyond the simple top-$M$ selection used here, and its relation to the accuracy-efficiency trade-off of the multi-path approach. Finally, scaling the benchmarks to larger qubit counts and deeper circuits, along with investigating the interplay between entanglement structure and tensor-network geometry in dynamic settings, are important steps toward applying this framework to practically relevant quantum circuits.

\section*{Acknowledgment}
The research is part of the Munich Quantum Valley (MQV), which is supported by the Bavarian state government with funds from the Hightech Agenda Bayern Plus.
A. Poggiali receives support by the Italian Project Fondo Italiano per la Scienza FIS00001966 ``MIMOSA'' and by the INdAM - GNCS project CUP E53C25002010001.

\bibliographystyle{IEEEtran}
\bibliography{biblio}

\end{document}